\documentclass[]{pasj02}
\begin{document} 
\Received{}
\Accepted{}
\title{Galactic absorption measured by X-ray observations of clusters of galaxies at the low Galactic latitude}

\author{
Yumiko \textsc{Anraku}\altaffilmark{1}, 
Shigeo \textsc{Yamauchi}\altaffilmark{1, $\ast$}, 
 Anje \textsc{Yoshimoto}\altaffilmark{1}, 
Masayoshi \textsc{Nobukawa}\altaffilmark{2}, 
Kumiko K. \textsc{Nobukawa}\altaffilmark{3}, 
and Hideki \textsc{Uchiyama}\altaffilmark{4}
}
\altaffiltext{1}{Faculty of Science, Nara Women's University, Kitauoyanishimachi, Nara, Nara 630-8506, Japan}
\altaffiltext{2}{Faculty of Education, Nara University of Education, Takabatake-cho, Nara, Nara 630-8528, Japan}
\altaffiltext{3}{Faculty of Science and Engineering, Kindai University, 3-4-1 Kowakae, Higashi-Osaka 577-8502, Japan}
\altaffiltext{4}{Faculty of Education, Shizuoka University, 836 Ohya, Suruga-ku, Shizuoka, Shizuoka 422-8529, Japan}
\email{yamauchi@cc.nara-wu.ac.jp}
\maketitle

\begin{abstract}
The amount of the interstellar gas in the Galaxy has been conventionally estimated through observations at various wavelengths. 
The estimation of the total hydrogen column density ($N_{\rm H}$) depends on assumptions such as temperature. 
The X-ray absorption process is the photoelectric absorption, which depends on the number of atoms to encounter X-ray photons, 
and hence X-ray observations would be able to derive the $N_{\rm H}$ values independently on the condition of the interstellar matter. 
We measured the Galactic absorption using clusters of galaxies at the low Galactic latitude. 
Comparing the observed $N_{\rm H}$ with the calculated $N_{\rm H}$ values from H\emissiontype{I} and CO intensities indicates
that the observed values are systematically larger than the calculated values. 
The observed $N_{\rm H}$ values at high Galactic latitude ($N_{\rm H}$ $< 10^{22}$ cm$^{-2}$) are comparable to those 
estimated from $N_{\rm H\emissiontype{I}}$ and optical reddening values using the method by Willingale et al. (2013, MNRAS, 431, 394),
but the values near to the Galactic plane ($N_{\rm H}$ $>10^{22}$ cm$^{-2}$) are larger than the estimated ones. 
The dust optical depth at 353 GHz, $\tau_{353}$, and the observed $N_{\rm H}$ values are 
expressed by a linear function of $N_{\rm H}=(1.01-1.59)\times 10^{26}\ \tau_{353}$ cm$^{-2}$ even at $N_{\rm H}$ $>10^{23}$ cm$^{-2}$.  
We also 
confirmed a linear correlation between the optical reddening, $E(B-V)$, and the $N_{\rm H}$ values expressed by $N_{\rm H}=(6.3-9.5)\times10^{21}\ E(B-V)\ {\rm cm}^{-2}$. 
This work is an additional and independent test of the relation among the amount of interstellar gas, the optical depth, and the optical reddening. 

\end{abstract}

\section{Introduction}

The Galaxy consists of stars, star clusters, interstellar matter (ISM), cosmic rays, and so on. 
ISM in the Galaxy is composed of gas and dust. 
ISM have been measured by survey observations in various wavelengths 
(e.g., \cite{DL1990,Dame2001,Schlegel1998,PlanckXXIV}). 

The amount of the interstellar gas in the Galaxy has been conventionally estimated by radio observations of 
atomic hydrogen (H\emissiontype{I}) and molecular hydrogen (H$_2$). 
The distribution of H\emissiontype{I} gas is observed by survey observations in the radio band at $\lambda$=21 cm (\cite{DL1990,Kalberla2005,HI4PI2016}). 
In the optically thin case, the H\emissiontype{I} column density ($N_{\rm H\emissiontype{I}}$) is expressed as follows \citep{DL1990}, 
\begin{equation}
N_{\rm H\emissiontype{I}} = 1.823\times10^{18} \times W_{\rm H\emissiontype{I}}, \\
\end{equation}
where $W_{\rm H\emissiontype{I}}$ is a velocity-integrated intensity of the H\emissiontype{I} emission at $\lambda$=21 cm. 

The amount of H$_2$ is estimated from CO line surveys at $\lambda$=2.6 mm (e.g., \cite{Dame2001}). 
The H$_2$ column density ($N_{\rm H_2}$) is expressed as follows \citep{Bolatto2013}, 
\begin{equation}
N_{\rm H_2} = X_{\rm CO} \times W_{\rm CO}, \\
\end{equation}
where $X_{\rm CO}$ is a conversion factor and $W_{\rm CO}$ is a velocity-integrated intensity of $^{12}$CO $J$=1--0. 
Based on the line observations in the radio band, 
the hydrogen column density, $N_{\rm H}$, is usually estimated as a sum of the column densities of H\emissiontype{I} and H$_{2}$, 
\begin{equation}
N_{\rm H}=N_{\rm H\emissiontype{I}} + 2 N_{\rm H_2}.
\end{equation}

The amount of dust is usually measured from extinction, reddening, or thermal emission at sub-millimeter to infrared wavelengths. 
The relation between optical extinction, $A_{\rm V}$, and $N_{\rm H}$ values has been observationally examined. 
Using two X-ray binaries together with the two extended sources, 
\citet{Reina1973} derived the first linear relation by  $N_{\rm H}=1.85 \times 10^{21} A_{\rm V}\ {\rm cm}^{-2}$. 
Then, 
several authors measured the relation using various objects, 
$N_{\rm H}=(1.79-2.22) \times 10^{21} A_{\rm V}\ {\rm cm}^{-2}$ \citep{Gorenstein1975,Predehl1995,Guever2009}. 
Recently, \citet{Gatuzz2024} reported the best-fit relation by $N_{\rm H} = (4.82\pm0.08)\times10^{21} A_{\rm V}+(0.59\pm0.16)\times10^{21}$ cm$^{-2}$ 
using coronal sources observed during the initial eROSITA all-sky survey. 
The relation between optical reddening, $E(B-V)$, and $N_{\rm H}$ values has also been examined. 
The relation was expressed by $N_{\rm H}=5.8\times10^{21}\ E(B-V)$ cm$^{-2}$ (e.g., \cite{Bohlin1978}) and 
recent studies proposed $N_{\rm H}=(7.7-9.4)\times10^{21}\ E(B-V)$ cm$^{-2}$
(e.g., \cite{Hensley2017, Lenz2017,Li2018}).

By comparing distributions of Gamma-rays and interstellar gas measured with H\emissiontype{I} and CO surveys, 
the amount of gas estimated from the Gamma-rays is larger than that estimated from H\emissiontype{I} and CO lines. 
Based on the results, a hypothesis that 
a considerable amount of gas is not properly traced by H\emissiontype{I} or CO surveys is proposed, called a dark gas scenario \citep{Grenier2005}. 
Planck observations have provided high-quality all-sky data at submillimeter wavelengths, including whole-sky distributions
of dust optical depth at 353 GHz ($\tau_{353}$) and dust temperature ($T_{\rm d}$) \citep{PlanckXI,PlanckX} and 
the dark gas was also suggested \citep{PlanckXXIV}. 
To explain the features, two hypotheses have been proposed, CO-dark H$_{2}$ gas (e.g., \cite{Wolfire2010,Smith2014}) and optically thick H\emissiontype{I} gas 
(e.g., \cite{Fukui2014}, 2015). 
Based on the good correlation between gas and dust distributions (e.g., \cite{Bohlin1978}), 
the total $N_{\rm H}$ value of ISM can be estimated from dust properties.
Assuming a constant gas-to-dust ratio and uniform dust properties, 
Fukui et al. (2014, 2015) examined a relationship between the $N_{\rm H\emissiontype{I}}$ value and $\tau_{353}$. 
They suggested that a relationship for the optically thin H\emissiontype{I} gas is good, but $W_{\rm H\emissiontype{I}}$ does not follow $\tau_{353}$ for the optically thick case: 
$W_{\rm CO}$ is saturated in the high density case on the Galactic plane. 
Thus, it is difficult to accurately assess the total $N_{\rm H}$ value from H\emissiontype{I} and CO surveys in high-density regions.

X-rays coming along lines of sight through the Galaxy is suffered from the photoelectric absorption by the ISM,  
but in the hard X-ray band, the interstellar medium is transparent even for the  line of sight through the Galactic plane. 
Therefore, X-ray observation for extragalactic objects 
in the wide energy band, such as 1--10 keV,  
is a powerful tool to measure the $N_{\rm H}$ value in the Galaxy. 

Using X-ray 
afterglows 
of gamma-ray bursts, energetic phenomena occurred in 
external galaxies, 
\citet{Willingale2013} investigated the total $N_{\rm H}$ value in the direction of the gamma-ray bursts. 
They derived a simple function for the $N_{\rm H_2}$ value 
composed of $N_{\rm H_I}$ and $E$($B-V$) as follows,  
\begin{equation}
N_{\rm H_2}=N_{\rm H_2 max} [1-{\rm exp}(\frac{-N_{\rm H\emissiontype{I}}E(B-V)}{N_{\rm c}})], 
\end{equation}
where $N_{\rm H_2 max}$ and $N_{\rm c}$ are constants to be derived from the fitting. 
Combining the $N_{\rm H\emissiontype{I}}$ value, 
they concluded that the empirical function represents a significant revision in Galactic absorption compared to the previous standard methods, particularly at low Galactic latitudes. 
However, two issues are considered. 
First, the estimation has uncertainty due to intrinsic absorption by the host galaxies where the Gamma-ray burst sources exist. 
Second, objects on the Galactic plane with a large $N_{\rm H}$ value ($>$10$^{22}$ cm$^{-2}$) are limited in their sample. 
Thus, a method to estimate the total $N_{\rm H}$ value 
on the Galactic plane of $b\sim0^{\circ}$ ($N_{\rm H}>10^{22}$ cm$^{-2}$)
has not yet been accurately established.  

Clusters of galaxies are bright X-ray sources with a luminosity of 10$^{40}$--10$^{46}$ erg s$^{-1}$ in the X-ray band (e.g., \cite{Fukazawa2004}). 
X-rays from clusters of galaxies originate from a hot plasma gas with a temperature of $\sim$0.5--10 keV 
and hence their spectra are simply expressed by a redshifted thin thermal plasma model. 
Since clusters of galaxies have no intrinsic absorption,  
they are more suitable target for investigating $N_{\rm H}$ of ISM than 
afterglows 
of gamma-ray bursts. 
Although the number of clusters of galaxies located near to the Galactic plane is limited,  
X-ray observations have discovered several candidates in the direction of low Galactic latitude  
(e.g., \cite{Nevalainen2001,LopesdeOliveira2006,Yamauchi2010,Yamauchi2011,Mori2013,Barriere2015, Nobukawa2015, Yamauchi2025a}). 
Some of them are located at $b\sim0^{\circ}$ and have large $N_{\rm H}$ values of 10$^{22}$--10$^{23}$ cm$^{-2}$. 
Thus, these are good samples to evaluate the amount of the ISM along the line of sight in the Galaxy. 

Based on the idea, we selected 17 spectral data of 16 clusters of galaxies and candidates located at the low Galactic latitude 
whose results have been published, and carried out reanalysis of the data. 
We compared the observed $N_{\rm H}$ values with values estimated from equation (3) and the method using equation (4) \citep{Willingale2013}.  
In addition, we also examined a relationship among the observed $N_{\rm H}$ value, $\tau_{353}$, $E(B-V)$ values, and 
confirmed that they are well correlated 
even at $N_{\rm H}>10^{23}$ cm$^{-2}$.
In this paper, we represent the results of the spectral analysis and the relationship. 

\section{X-ray data}

%
\begin{table*}[t]
\caption{Observation logs.}
\begin{center} \label{tab:log}
\begin{tabular}{lcccc} \hline  
Target name  & Obs. position RA, Dec (J2000.0)& Start time & Exposure (ks) & Note$^{\ast}$\\ 
Observation ID&  $l, b$ & End time  & & \\ \hline
\multicolumn{5}{c}{ASCA} \\ 
AX J145732$-$5901			& \timeform{14h58m30s.61}, \timeform{-58D55'32''.0}	& 1999-02-23 23:40:34	& 11.2 & 1\\
57004070					& \timeform{318D.700}, \timeform{-0D.000}			& 1999-02-24 10:30:41	& & \\ 

3C129.1					& \timeform{4h49m35s.47 }, \timeform{+45D06'43''.0}	& 1998-08-31 01:34:41	& 44.1 &2 \\
86050000					& \timeform{160D.400}, \timeform{+0D.266}			& 1998-09-01 12:10:41	& &\\
\hline
\multicolumn{5}{c}{Suzaku} \\ 
AX J185905$+$0333		&  \timeform{18h59m12s.22}, \timeform {+3D34'52''.7} 	& 2010-04-17 17:35:13 	&  51.0 & 3\\
505027010  				& \timeform{37D.004}, \timeform{-0D.092} 			& 2010-04-18 21:09:18 	& &\\
 
Suzaku J1840.2$-$0544		& \timeform{18h40m35s.30}, \timeform{-5D36'32''.4}		& 2011-03-27 15:21:16	& 49.6 &4\\
505090010				& \timeform{26D.705}, \timeform{-0D.152}				& 2011-03-28 19:00:18	& &\\
 
CIZA J2242.8$+$5301		&  \timeform{22h43m02s.45},  \timeform{+53D09'41''.8}	& 2011-07-28 13:19:44 	& 122.9 &5\\
806001010				&  \timeform{104D.280},  \timeform{-4D.996} 			& 2011-07-30 19:42:16	& &\\
 
Suzaku J1759$-$3450		&  \timeform{17h59m10s.13},  \timeform{-34D49'09''.8} 	& 2012-03-07 21:40:15 	& 40.2 &6\\
406019010				&  \timeform{356D.383},  \timeform{-5D.461} 			& 2012-03-08 21:54:13 	& &\\
 
Cygnus A Cluster			&  \timeform{19h59m02s.62},  \timeform{+40D47'18''.6} 	& 2008-11-15 21:43:08 	& 44.7 &7\\
803050010				&  \timeform{76D.195},  \timeform{+5D.853} 			& 2008-11-16 21:58:19 	& &\\
 
2XMM J045637.2$+$522411 	&  \timeform{04h56m53s.52},  \timeform{+52D25'02''.6} 	& 2007-02-16 15:40:27	& 50.5 &8\\
501075010				&  \timeform{155D.484},  \timeform{+5D.809} 			& 2007-02-17 18:08:24 	& &\\
 
CIZA J1700.84$-$3144 		&  \timeform{17h 00m 47s .04},  \timeform{-31D44'42''.7} 	& 2013-02-18 11:50:13 	& 8.9 &9\\
407027010				&  \timeform{352D.212},  \timeform{+6D.400} 			& 2013-02-18 18:00:11 	& &\\
 
A3627 					&  \timeform{16h14m16s.13},  \timeform{-60D50'59''.6} 	& 2009-03-15 22:09:21 	& 46.0 &10\\
803032010				&  \timeform{325D.259},  \timeform{-7D.106} 			& 2009-03-16 14:34:24 	& &\\
 
Ophiuchus Cluster	 		&  \timeform{17h12m26s.23},  \timeform{-23D22'44''.4} 	& 2007-09-24 21:12:15 	& 100.5 &11\\
802046010				&  \timeform{0D.562},  \timeform{+9D.269} 			& 2007-09-27 14:15:16 	& &\\
 
CIZA J1358.9$-$4750 		&  \timeform{13h58m36s.24},  \timeform{-47D46'32''.9} 	& 2013-01-21 18:07:12 	& 61.7 &12\\
807037010				&  \timeform{314D.453},  \timeform{+13D.588} 			& 2013-01-23 11:36:20 	& &\\
\hline 
\multicolumn{5}{c}{XMM-Newton} \\ 
XMMU J183225.4$-$103645 	&  \timeform{18h32m51s.97},  \timeform{-10D35'47''.0} 	& 2000-04-11 12:26:55 	& 34.2 &13\\
0122700301				&  \timeform{21D.397},  \timeform{-0D.747} 			& 2000-04-11 21:57:00 	& &\\
 
IGR J17448$-$3232 			&  \timeform{17h44m53s.41},  \timeform{-32D32'54''.0} 	& 2012-02-26 04:16:42 	& 43.9 &14\\
0672260101				&  \timeform{356D.836},  \timeform{-1D.750} 			& 2012-02-26 16:28:41 	& &\\
 
CIZA J1324.7$-$5736 		&  \timeform{13h24m47s.02},  \timeform{-57D36'34''.0} 	& 2003-08-16 20:49:50 	& 12.2 &15\\
0152131201				&  \timeform{307D.403},  \timeform{+4D.973} 			& 2003-08-17 00:13:58 	& &\\
 
Cl 2334$+$48 				&  \timeform{23h33m39s.96},  \timeform{+48D49'05''.9} 	& 2001-01-28 15:00:38 	& 25.2 &16\\
0093552701				&  \timeform{109D.979},  \timeform{-12D.088} 			& 2001-01-28 22:01:00 	& &\\
\hline\\
\end{tabular}
\end{center}
\vspace{-10pt}
$^{\ast}$ Previous works. 1: \citet{Yamauchi2025a}, 2: \citet{Taylor2001}, 3: \citet{Yamauchi2011}, 4: \citet{Nobukawa2015}, 5: \citet{Akamatsu2013}, 6 : \citet{Mori2013},  7: \citet{Smith2002}, 8: \citet{Yamauchi2010}, 
9: \citet{Mori2017}, 10: \citet{Nishino2012}, 
11: \citet{Fujita2008}, 12: \citet{Kato2015}, 13: \citet{Nevalainen2001}, 14: \citet{Barriere2015}, 15: \citet{Mullis2005}, 16: \citet{LopesdeOliveira2006}. 
\end{table*} 

We utilized the ASCA, Suzaku, and XMM-Newton archival data. 
The ASCA and Suzaku data were provided by DARTS/JAXA, whereas 
the XMM-Newton data were provided by XMM-Newton Science Archive/ESA. 

The ASCA observations were carried out with the two Solid-state Imaging 
Spectrometers (SIS0, SIS1:  \cite{Burke1991}) and the two Gas Imaging Spectrometers (GIS2, GIS3:  \cite{Makishima1996, Ohashi1996}) 
placed at the focal plane of the thin foil X-ray Telescope (XRT: \cite{Serlemitsos1995}). 
Since the GIS had a large field of view to cover the whole target, we utilized the GIS data. 
The GIS was operated in PH mode. 
We analyzed data of a cluster of galaxies and a candidate,  
3C129.1 and AX J145732$-$5901. 

The Suzaku data were obtained from the X-ray Imaging Spectrometer (XIS: \cite{Ko07}) 
placed at the focal planes of the thin foil X-ray Telescopes (XRT: \cite{Se07}). 
The XIS was operated in the normal clocking mode.
We analyzed 11 spectral data of 10 clusters of galaxies and candidates,  
AX J185905$+$0333, Suzaku J1840.2$-$0544, CIZA 2242.8$+$5301, Suzaku J1759$-$3450, Cygnus A Cluster, 
2XMM J045637.2$+$522411,  CIZA J1700.84$-$3144, A3627, 
Ophiuchus Cluster, 
northwest (NW) and southeast (SE) regions of CIZA J1358.9$-$4750. 

The XMM-Newton data were obtained with the European Photon Imaging Camera (EPIC).
The EPIC system is composed of two different detectors; 
two MOS cameras (MOS 1 and MOS 2: \cite{Turner2001}) and one pn camera \citep{Struder2001}. 
The MOS and pn cameras were operated in the full-frame mode. 
We analyzed data of 
4 clusters of galaxies and candidates, 
XMMU J183225.4$-$103645, IGR J17448$-$3232, 
CIZA J1324.7$-$5736, and Cl 2334$+$48. 
The logs of the ASCA, Suzaku, and XMM-Newton observations are listed in table \ref{tab:log}.

\section{Analysis and results}
\subsection{Spectral analysis}

We conducted spectral analysis for ASCA, Suzaku, and XMM-Newton data. 
The ASCA and Suzaku data analysis was performed using the HEAsoft, 
whereas the XMM-Newton data analysis was performed using the Extended Source Analysis Software (ESAS) package and the HEAsoft. 

A source spectrum was extracted from a source region centered on the X-ray emission, 
whereas a background spectrum was estimated from nearby blank sky data.  
According to \citet{Snowden2008}, for the XMM-Newton data, we took account of soft proton contamination 
(a time-variable flare component and a quiescent continuum component) and instrumental fluorescence lines 
for the non-X-ray background.  

The source spectrum after excluding the sky and non-X-ray background components 
was fitted with a thin thermal plasma model ({\tt apec} in XSPEC, \cite{Smith2001}) 
modified by 
photoelectric absorption. 
The atomic data of the lines and continua of the thin thermal plasma were taken from ATOMDB 3.0.9. 
For the abundance table, several data are available. 
Here, we utilized three abundance tables from \citet{Anders1989} (hereafter {\tt angr}), 
\citet{Lodders2003} (hereafter {\tt lodd}), and \citet{Wilms2000} (hereafter {\tt wilm}). 
For the model of 
the photoelectric absorption, 
we utilized a {\tt phabs} model 
with the cross-sections of the photoelectric absorption taken from \citet{Vern}.
In this analysis, we assumed the abundances of ISM to be solar. 
Thus, we applied the spectral model as follows, 

\smallskip

Model 1 {\tt (phabs, angr)}: {\tt phabs$\times$apec} ({\tt angr})

\smallskip

Model 2  {\tt (phabs, lodd)}: {\tt phabs$\times$apec} ({\tt lodd})

\smallskip

Model 3  {\tt (phabs, wilm)}: {\tt phabs$\times$apec} ({\tt wilm})

\smallskip

To examine the difference in the cross sections of low-energy absorption, 
we utilized another absorption model, {\tt tbabs} \citep{Wilms2000}
which calculates the cross sections for X-ray absorption due to the gas-phase ISM, the grain-phase ISM, and the molecules in the ISM. 
The applied spectral model is as follows, 

\smallskip

Model 4  {\tt (tbabs, wllm)}: {\tt tbabs$\times$apec} ({\tt wilm})

\smallskip 

Free parameters are normalization, electron temperature, redshift, and abundance of the plasma gas in clusters of galaxies, and the $N_{\rm H}$ value of photoelectric absorption. 
The best-fitting results are listed in tables \ref{tab:fit1}-\ref{tab:fit3} for each target. 
We note that the results are basically consistent with those in the previous works listed in tables \ref{tab:fit1}-\ref{tab:fit3}. 

Comparing the results obtained using three abundance tables,  
we found that the $N_{\rm H}$ values using {\tt lodd} and {\tt wilm} are consistent with each other, 
but that using {\tt angr} is systematically smaller than the others. 
On the other hand, the results using the {\tt phabs} model (model 3) and the {\tt tbabs} model (model 4) are almost the same. 

%
\begin{table*}[t]
\caption{Best-fitting parameters for each target.}
\small
\begin{center}\label{tab:fit1}
\begin{tabular}{lcccc} \hline 
Parameter & Model 1 & Model 2 & Model 3 & Model 4 \\ 
                  & {\tt (phabs,  angr)}     & {\tt (phabs, lodd)}     & {\tt (phabs, wilm)}       &{\tt (tbabs, wilm)} \\ 
\hline
& \multicolumn{4}{c}{AX J145732$-$5901} \\ 
$N_{\rm H}$ (10$^{22}$ cm$^{-2}$) 		& 13$^{+15}_{-7}$  		& 18$^{+22}_{-9}$		&19$^{+24}_{-10}$		& 19$^{+23}_{-10}$ \\
$kT_{\rm e}$ (keV) 					& 2.6$^{+4.9}_{-1.5}$  	& 2.6$^{+4.6}_{-1.5}$	& 2.7$^{+4.8}_{-1.5}$	&2.7$^{+4.8}_{-1.5}$\\
 Abundance (Solar)					& $>$0.36		 		& $>$0.52				& $>$0.63				& $>$0.64\\
Redshift							& 0.12$\pm$0.04		& 0.12$\pm$0.04		& 0.12$\pm$0.04		& 0.13$\pm$0.04 \\
Normalization$^{\ast}$ (10$^{-3}$ cm$^{-5}$) 		& 8.2$^{+67.0}_{-6.4}$	& 8.6$^{+74.5}_{-6.7}$	&7.5$^{+66.9}_{-5.7}$ 	& 7.7$^{+66.3}_{-7.0}$ \\
 $\chi^2$/d.o.f. 						&  27.5/31				&27.8/31				& 27.7/31				& 27.7/31	 \\
 \hline
& \multicolumn{4}{c}{AX J185905$+$0333} \\
$N_{\rm H}$ (10$^{22}$ cm$^{-2}$) 		& 10.6$\pm$1.4  		& 15.0$^{+2.1}_{-1.6}$	&16.5$^{+2.3}_{-1.8}$ 	& 16.2$^{+2.2}_{-1.8}$ \\
$kT_{\rm e}$ (keV) 					& 5.8$\pm$1.1 	 		& 5.7$\pm$1.1			& 5.8$\pm$1.1			& 5.8$\pm$1.1\\
 Abundance (Solar) 					& 0.32$\pm$0.12 		& 0.47$\pm$0.18		& 0.55$^{+0.21}_{-0.16}$	& 0.55$^{+0.22}_{-0.16}$\\
Redshift							& 0.393$\pm$0.006		& 0.393$\pm$0.006		& 0.393$\pm$0.006		& 0.393$\pm$0.006 \\
Normalization$^{\ast}$ (10$^{-3}$ cm$^{-5}$)  						& 5.6$\pm$1.9			& 6.2$\pm$2.2			& 5.6$\pm$2.0			& 5.6$\pm$2.0\\
 $\chi^2$/d.o.f. 						&  116.8/136			&116.0/136 			& 116.6/136			& 116.6/136 \\
\hline
& \multicolumn{4}{c}{Suzaku J1840.2$-$0544} \\
$N_{\rm H}$ (10$^{22}$ cm$^{-2}$) 		& 5.6$\pm$1.1   		& 8.2$\pm$1.6			&8.8$\pm$1.8 			& 8.6$\pm$1.8 \\
$kT_{\rm e}$ (keV) 					& 6.2$\pm$1.5 	 		& 6.1$\pm$1.5			& 6.2$\pm$1.4			& 6.2$\pm$1.5\\
 Abundance (Solar)  					& 0.61$^{+0.23}_{-0.18}$ 	& 0.9$\pm$0.4			& 1.0$\pm$0.4			& 1.0$\pm$0.4\\
Redshift							& 0.093$\pm$0.007		& 0.093$\pm$0.007		& 0.093$\pm$0.007		& 0.093$\pm$0.007 \\
Normalization$^{\ast}$ (10$^{-3}$ cm$^{-5}$)  						& 2.3$\pm$0.6			&2.5$\pm$0.7 			& 2.3$\pm$0.7			& 2.3$\pm$0.7 \\
 $\chi^2$/d.o.f. 						&  112.9/104			&112.4/104 			& 113.0/104			& 113.1/104 \\
\hline
& \multicolumn{4}{c}{3C129.1} \\
$N_{\rm H}$ (10$^{22}$ cm$^{-2}$) 		& 0.70$\pm$0.03   		& 1.08$\pm$0.04		&1.03$\pm$0.04 		& 1.01$\pm$0.04 \\
$kT_{\rm e}$ (keV) 					& 6.4$\pm$0.3 	 		& 6.2$\pm$0.4			& 6.5$\pm$0.3			& 6.5$\pm$0.3\\
 Abundance (Solar)  					& 0.34$\pm$0.04	 	& 0.51$\pm$0.06		& 0.59$\pm$0.07		& 0.59$\pm$0.07\\
Redshift							& 0.029$^{+0.002}_{-0.009}$		& 0.025$\pm$0.007		& 0.029$^{+0.002}_{-0.009}$		& 0.028$^{+0.003}_{-0.008}$ \\
Normalization$^{\ast}$ (10$^{-3}$ cm$^{-5}$)   						& 76$\pm$2		& 81$\pm$2		& 75$\pm$2		& 75$\pm$2\\
 $\chi^2$/d.o.f. 						&  783.7/656			&784.6/656 			& 781.0/656			& 780.8/656 \\
\hline
& \multicolumn{4}{c}{XMMU J183225.4$-$103645} \\
$N_{\rm H}$ (10$^{22}$ cm$^{-2}$) 		&  7.9$\pm$0.9  		& 11.6$\pm$1.2			&12.4$\pm$1.3 		& 12.1$\pm$1.3\\
$kT_{\rm e}$ (keV) 					&  5.6$\pm$1.1		 	& 5.4$\pm$1.1			& 5.6$\pm$1.1			& 5.6$\pm$1.1\\
 Abundance (Solar) 					& 0.60$\pm$0.18		& 0.85$\pm$0.25		& 1.03$\pm$0.31		& 1.03$\pm$0.31\\
Redshift							& 0.125$\pm$0.004	 	& 0.125$\pm$0.005		&0.125$\pm$0.004		& 0.125$\pm$0.004 \\
Normalization$^{\ast}$ (10$^{-3}$ cm$^{-5}$)  						& 2.6$\pm$0.6			& 2.9$\pm$0.7			& 2.6$\pm$0.6			& 2.6$\pm$0.6\\
 $\chi^2$/d.o.f. 						& 122.1/115			&121.6/115 			& 122.4/115			& 122.4/115 \\
\hline
& \multicolumn{4}{c}{IGR J17448$-$3232 	} \\
$N_{\rm H}$ (10$^{22}$ cm$^{-2}$) 		& 2.21$\pm$0.07 		& 3.34$\pm$0.09 		&3.35$\pm$0.08  		& 3.26$\pm$0.09 \\
$kT_{\rm e}$ (keV) 					& 11.1$\pm$0.9		 	& 10.3$\pm$0.7		&11.6$\pm$1.0			& 11.6$\pm$1.0\\
 Abundance (Solar) 					&  0.31$\pm$0.04		& 0.44$\pm$0.05		& 0.57$\pm$0.07		& 0.57$\pm$0.07\\
Redshift							& 0.054$\pm$0.002	 	& 0.054$\pm$0.002		& 0.054$\pm$0.003		& 0.054$\pm$0.003 \\
Normalization$^{\ast}$ (10$^{-3}$ cm$^{-5}$) 						& 29.1$\pm$0.6		& 32.0$\pm$0.7		& 28.6$\pm$0.5		& 28.5$\pm$0.5\\
 $\chi^2$/d.o.f. 						& 1405.5/1496			&1377.3/1496			& 1427.4/1496			& 1429.3/1496 \\
 \hline 
\\
\end{tabular}
\end{center}
\vspace{-10pt}
Errors are provided at 90\% confidence level.\\
$^{\ast}$ Defined as 10$^{-14}$ $\times$$\int n_{\rm H} n_{\rm e} dV$/[4$\pi D_{\rm A}^2(1+z)^2$],
where 
$n_{\rm H}$ is the hydrogen density (cm$^{-3}$), 
$n_{\rm e}$ is the electron density (cm$^{-3}$), 
$D_{\rm A}$ is the angular distance (cm), $z$ is the redshift, and $V$ is the volume (cm$^3$). \\
\\
\normalsize
\end{table*}


%
\begin{table*}[t]
\caption{Best-fitting parameters for each target.}
\small
\begin{center}\label{tab:fit2}
\begin{tabular}{lcccc} \hline 
Parameter & Model 1 & Model 2 & Model 3  & Model 4 \\ 
                  &{\tt (phabs,  angr)}       & {\tt (phabs, lodd)}      & {\tt (phabs, wilm)}      & {\tt (tbabs, wilm)}\\ 
\hline
& \multicolumn{4}{c}{CIZA J1324.7$-$5736} \\
$N_{\rm H}$ (10$^{22}$ cm$^{-2}$) 		& 0.54$\pm$0.02 		& 0.76$\pm$0.03 		& 0.75$\pm$0.03		&  0.72$\pm$0.03 \\
$kT_{\rm e}$ (keV) 					& 2.42$\pm$0.06	 	& 2.47$\pm$0.07		& 2.47$\pm$0.07		& 2.49$\pm$0.07\\
 Abundance (Solar) 					& 0.61$\pm$0.06		& 0.88$\pm$0.10		& 1.04$\pm$0.11		& 1.03$\pm$0.11\\
Redshift							& 0.019$\pm$0.002	 	& 0.019$\pm$0.004		& 0.019$\pm$0.002		& 0.019$\pm$0.002 \\
Normalization$^{\ast}$ (10$^{-3}$ cm$^{-5}$)  						& 13.6$\pm$0.6		& 13.6$\pm$0.6		& 13.1$\pm$0.6		& 13.0$\pm$0.6\\
 $\chi^2$/d.o.f. 						& 1016.2/1047			&997.8/1047 			& 1009.1/1047			& 1007.5/1047\\
\hline
 & \multicolumn{4}{c}{CIZA J2242.8$+$5301} \\
$N_{\rm H}$ (10$^{22}$ cm$^{-2}$) 		& 0.36$\pm$0.02    		& 0.55$\pm$0.03		&0.52$\pm$0.03 		& 0.51$\pm$0.03 \\
$kT_{\rm e}$ (keV) 					& 7.4$\pm$0.5 	 		& 7.4$\pm$0.5			& 7.5$\pm$0.5			& 7.6$\pm$0.5\\
 Abundance (Solar)  					& 0.22$\pm$0.05  		& 0.33$\pm$0.07		& 0.39$\pm$0.08		& 0.39$\pm$0.08\\
Redshift							& 0.190$\pm$0.005		& 0.190$\pm$0.006		& 0.190$\pm$0.006		& 0.190$\pm$0.006 \\
Normalization$^{\ast}$ (10$^{-3}$ cm$^{-5}$)  						& 3.71$\pm$0.10		& 3.91$\pm$0.11		& 3.65$\pm$0.10		& 3.64$\pm$0.10 \\
 $\chi^2$/d.o.f. 						& 274.0/262			&271.1/262 			& 274.7/262			& 275.0/262 \\
\hline
 & \multicolumn{4}{c}{Suzaku J1759$-$3450} \\
$N_{\rm H}$ (10$^{22}$ cm$^{-2}$) 		& 0.31$\pm$0.04    		& 0.47$\pm$0.05 		&0.45$\pm$0.05 		& 0.43$\pm$0.05  \\
$kT_{\rm e}$ (keV) 					& 5.0$\pm$0.4 	 		& 5.0$\pm$0.4			& 5.0$\pm$0.4			& 5.0$\pm$0.4\\
 Abundance (Solar)					& 0.15$\pm$0.05  		& 0.23$\pm$0.08		& 0.27$\pm$0.09		&  0.27$\pm$0.09\\
Redshift							&  0.130$\pm$0.007		& 0.130$\pm$0.007		& 0.130$\pm$0.007		& 0.130$\pm$0.007 \\
Normalization$^{\ast}$ (10$^{-3}$ cm$^{-5}$)  						& 4.6$\pm$0.2			& 4.8$\pm$0.3			& 4.5$\pm$0.2			& 4.5$\pm$0.2	\\
 $\chi^2$/d.o.f. 						& 195.7/172			&195.1/172 			& 196.2/172			& 196.7/172 \\
\hline
  & \multicolumn{4}{c}{Cygnus A Cluster} \\
$N_{\rm H}$ (10$^{22}$ cm$^{-2}$) 		& 0.287$\pm$0.011   	& 0.415$\pm$0.011  		&0.401$\pm$0.010  		& 0.385$\pm$0.010 \\
$kT_{\rm e}$ (keV) 					& 5.5$\pm$0.4 	 		& 5.8$\pm$0.3			& 5.8$\pm$0.3			& 5.9$\pm$0.3\\
 Abundance (Solar)  					& 0.49$\pm$0.04  		& 0.75$\pm$0.05		& 0.89$\pm$0.05		& 0.89$\pm$0.06\\
Redshift							&  0.058$\pm$0.001		& 0.058$\pm$0.001		& 0.058$\pm$0.001		& 0.058$\pm$0.001 \\
Normalization$^{\ast}$ (10$^{-3}$ cm$^{-5}$)  						& 38.5$\pm$0.7		& 39.9$\pm$0.5		& 37.6$\pm$0.5		&37.4$\pm$0.5 \\
 $\chi^2$/d.o.f. 						& 2658.9/2502			&2651.1/2502 			& 2661.2/2502			& 2668.0/2502 \\
\hline
& \multicolumn{4}{c}{2XMM J045637.2$+$522411} \\
$N_{\rm H}$ (10$^{22}$ cm$^{-2}$) 		& 0.34$\pm$0.20   		& 0.50$\pm$0.29 		& 0.49$\pm$0.28 		& 0.48$\pm$0.28 \\
$kT_{\rm e}$ (keV) 					& 3.8$\pm$1.0 	 		& 3.9$\pm$1.0			& 3.9$\pm$1.0			& 3.8$\pm$1.0 \\
 Abundance (Solar)  					& 0.27$\pm$0.22  		& 0.40$\pm$0.32		& 0.50$\pm$0.38		& 0.49$\pm$0.39\\
Redshift							&  0.162$\pm$0.016		& 0.163$\pm$0.017		&0.163$\pm$0.016		& 0.161$\pm$0.016 \\
Normalization$^{\ast}$ (10$^{-3}$ cm$^{-5}$) 						& 0.70$\pm$0.16		&0.72$\pm$0.16 		& 0.68$\pm$0.15		& 0.68$\pm$0.15\\
 $\chi^2$/d.o.f. 						& 92.5/105 			& 93.2/105			& 92.3/105			& 92.3/105 \\
\hline
 & \multicolumn{4}{c}{CIZA J1700.84$-$3144} \\
$N_{\rm H}$ (10$^{22}$ cm$^{-2}$) 		& 0.31$\pm$0.05   		& 0.46$\pm$0.07  		&0.45$\pm$0.07 		& 0.43$\pm$0.07  \\
$kT_{\rm e}$ (keV) 					& 5.4$\pm$0.7 	 		& 5.4$\pm$0.7			& 5.5$\pm$0.7			& 5.5$\pm$0.7\\
 Abundance (Solar)					& 0.29$^{+0.10}_{-0.05}$	& 0.44$\pm$0.14		& 0.51$\pm$0.17		& 0.51$\pm$0.16\\
Redshift							&  0.136$\pm$0.008		& 0.136$\pm$0.008		& 0.137$\pm$0.008		& 0.137$\pm$0.008 \\
Normalization$^{\ast}$ (10$^{-3}$ cm$^{-5}$)  						& 6.5$\pm$0.4			& 6.8$\pm$0.5			& 6.4$\pm$0.4			& 6.4$\pm$0.4\\
 $\chi^2$/d.o.f. 						& 177.9/169			&178.7/169 			& 179.3/169			& 180.2/169 \\
\hline \\
\end{tabular}
\end{center}
\vspace{-10pt}
Errors are provided at 90\% confidence level.\\
$^{\ast}$ Defined as 10$^{-14}$ $\times$$\int n_{\rm H} n_{\rm e} dV$/[4$\pi D_{\rm A}^2(1+z)^2$],
where 
$n_{\rm H}$ is the hydrogen density (cm$^{-3}$), 
$n_{\rm e}$ is the electron density (cm$^{-3}$), 
$D_{\rm A}$ is the angular distance (cm), $z$ is the redshift, and $V$ is the volume (cm$^3$). \\
\normalsize
\end{table*}

%
\begin{table*}[t]
\caption{Best-fitting parameters for each target.}
\small
\begin{center}\label{tab:fit3}
\begin{tabular}{lcccc} \hline 
Parameter & Model 1 & Model 2 & Model 3  & Model 4 \\ 
                  &{\tt (phabs,  angr)}       & {\tt (phabs, lodd)}      & {\tt (phabs, wilm)}      & {\tt (tbabs, wilm)}\\ 
\hline
 & \multicolumn{4}{c}{A3627} \\
$N_{\rm H}$ (10$^{22}$ cm$^{-2}$) 		& 0.219$\pm$0.012  		& 0.33$\pm$0.02		&0.32$\pm$0.02 		& 0.31$\pm$0.02  \\
$kT_{\rm e}$ (keV) 					& 6.0$\pm$0.2 	 		& 6.1$\pm$0.2			& 6.1$\pm$0.2			& 6.1$\pm$0.2\\
 Abundance (Solar) 					& 0.31$\pm$0.03		& 0.47$\pm$0.05		& 0.55$\pm$0.05		& 0.55$\pm$0.05\\
Redshift							&  0.017$\pm$0.002		& 0.017$\pm$0.002		& 0.017$\pm$0.002		& 0.017$\pm$0.002 \\
Normalization$^{\ast}$ (10$^{-3}$ cm$^{-5}$) 						& 11.3$\pm$0.2			& 11.9$\pm$0.2			& 11.2$\pm$0.2			& 11.2$\pm$0.3\\
 $\chi^2$/d.o.f. 						& 379.5/352			&378.3/352 			& 378.2/352			& 378.3/352 \\
\hline
& \multicolumn{4}{c}{Ophiuchus Cluster} \\
$N_{\rm H}$ (10$^{22}$ cm$^{-2}$) 		& 0.349$\pm$0.003  		& 0.519$\pm$0.005		& 0.500$\pm$0.004 		& 0.483$\pm$0.005 \\
$kT_{\rm e}$ (keV) 					& 8.08$\pm$0.07	 	& 8.14$\pm$0.08		&  8.25$\pm$0.07		& 8.28$\pm$0.07\\
 Abundance (Solar)  					& 0.427$\pm$0.011 		& 0.64$\pm$0.02		& 0.76$\pm$0.02		& 0.76$\pm$0.02\\
Redshift							& 0.0293$\pm$0.0003 	& 0.0293$\pm$0.0003	& 0.0293$\pm$0.0003	& 0.0293$\pm$0.0003 \\
Normalization$^{\ast}$ (10$^{-3}$ cm$^{-5}$)  						& 70.3$\pm$0.3		& 73.7$\pm$0.4		& 69.1$\pm$0.3		& 68.9$\pm$0.3\\
 $\chi^2$/d.o.f. 						& 3551.9/3115			&3454.9/3115 			& 3501.0/3115			& 3518.0/3115 \\
\hline
& \multicolumn{4}{c}{Cl 2334$+$48 } \\
$N_{\rm H}$ (10$^{22}$ cm$^{-2}$) 		& 0.19$\pm$0.02 		& 0.25$\pm$0.03		&0.24$\pm$0.03		& 0.22$\pm$0.03 \\
$kT_{\rm e}$ (keV) 					& 4.1$\pm$0.6	 		& 4.5$\pm$0.5			& 4.6$\pm$0.5			& 4.6$\pm$0.5\\
 Abundance (Solar) 					& 0.44$\pm$0.15		& 0.63$\pm$0.21		& 0.72$\pm$0.25		& 0.72$\pm$0.25\\
Redshift							&0.28$\pm$0.03	 	& 0.29$\pm$0.03		& 0.29$\pm$0.02		& 0.29$\pm$0.03 \\
Normalization$^{\ast}$ (10$^{-3}$ cm$^{-5}$)  						& 1.20$\pm$0.10		& 1.21$\pm$0.10		& 1.14$\pm$0.09		& 1.13$\pm$0.13\\
 $\chi^2$/d.o.f. 						& 235.6/268			&237.0/268 			& 238.5/268			& 238.7/268 \\
\hline
& \multicolumn{4}{c}{CIZA J1358.9$-$4750 (NW)} \\
$N_{\rm H}$ (10$^{22}$ cm$^{-2}$) 		&  0.17$\pm$0.02  		&  0.24$\pm$0.03 		& 0.23$\pm$0.03 		&  0.22$\pm$0.03  \\
$kT_{\rm e}$ (keV) 					&  3.6$\pm$0.2	 		& 3.7$\pm$0.2			& 3.7$\pm$0.2			& 3.7$\pm$0.2\\
 Abundance (Solar)  					& 0.26$\pm$0.06		& 0.39$\pm$0.09		& 0.45$\pm$0.11		& 0.45$\pm$0.11\\
Redshift							& 0.07$\pm$0.02	 	& 0.07$\pm$0.02		& 0.07$\pm$0.02		&0.07$\pm$0.03 \\
Normalization$^{\ast}$ (10$^{-3}$ cm$^{-5}$)  						& 4.1$\pm$0.3			& 4.2$\pm$0.3			& 4.0$\pm$0.3			& 4.0$\pm$0.3\\
 $\chi^2$/d.o.f. 						& 158.0/145			&153.5/145 			& 156.0/145			& 155.7/145 \\
\hline 
& \multicolumn{4}{c}{CIZA J1358.9$-$4750 (SE)} \\
$N_{\rm H}$ (10$^{22}$ cm$^{-2}$) 		&  0.13$\pm$0.02  		&  0.19$\pm$0.03  		&0.18$\pm$0.03  		& 0.18$\pm$0.02  \\
$kT_{\rm e}$ (keV) 					&  5.6$\pm$0.4		 	& 5.6$\pm$0.4			& 5.7$\pm$0.4			& 5.7$\pm$0.4\\
 Abundance (Solar) 					& 0.45$\pm$0.10		& 0.68$\pm$0.11		& 0.79$\pm$0.12		& 0.79$\pm$0.12\\
Redshift							& 0.079$\pm$0.006	 	& 0.079$\pm$0.006		& 0.079$\pm$0.009		& 0.079$\pm$0.007 \\
Normalization$^{\ast}$ (10$^{-3}$ cm$^{-5}$)  						& 4.9$\pm$0.2			& 5.1$\pm$0.2			& 4.8$\pm$0.2			& 4.8$\pm$0.2\\
 $\chi^2$/d.o.f. 						& 220.8/235			&224.5/235 			& 223.1/235			& 223.8/235 \\
\hline \\
\end{tabular}
\end{center}
\vspace{-10pt}
Errors are provided at 90\% confidence level.\\
$^{\ast}$ Defined as 10$^{-14}$ $\times$$\int n_{\rm H} n_{\rm e} dV$/[4$\pi D_{\rm A}^2(1+z)^2$],
where 
$n_{\rm H}$ is the hydrogen density (cm$^{-3}$), 
$n_{\rm e}$ is the electron density (cm$^{-3}$), 
$D_{\rm A}$ is the angular distance (cm), $z$ is the redshift, and $V$ is the volume (cm$^3$). \\
\normalsize
\end{table*}

\begin{figure*}
  \begin{center}
    \FigureFile(15cm,22cm){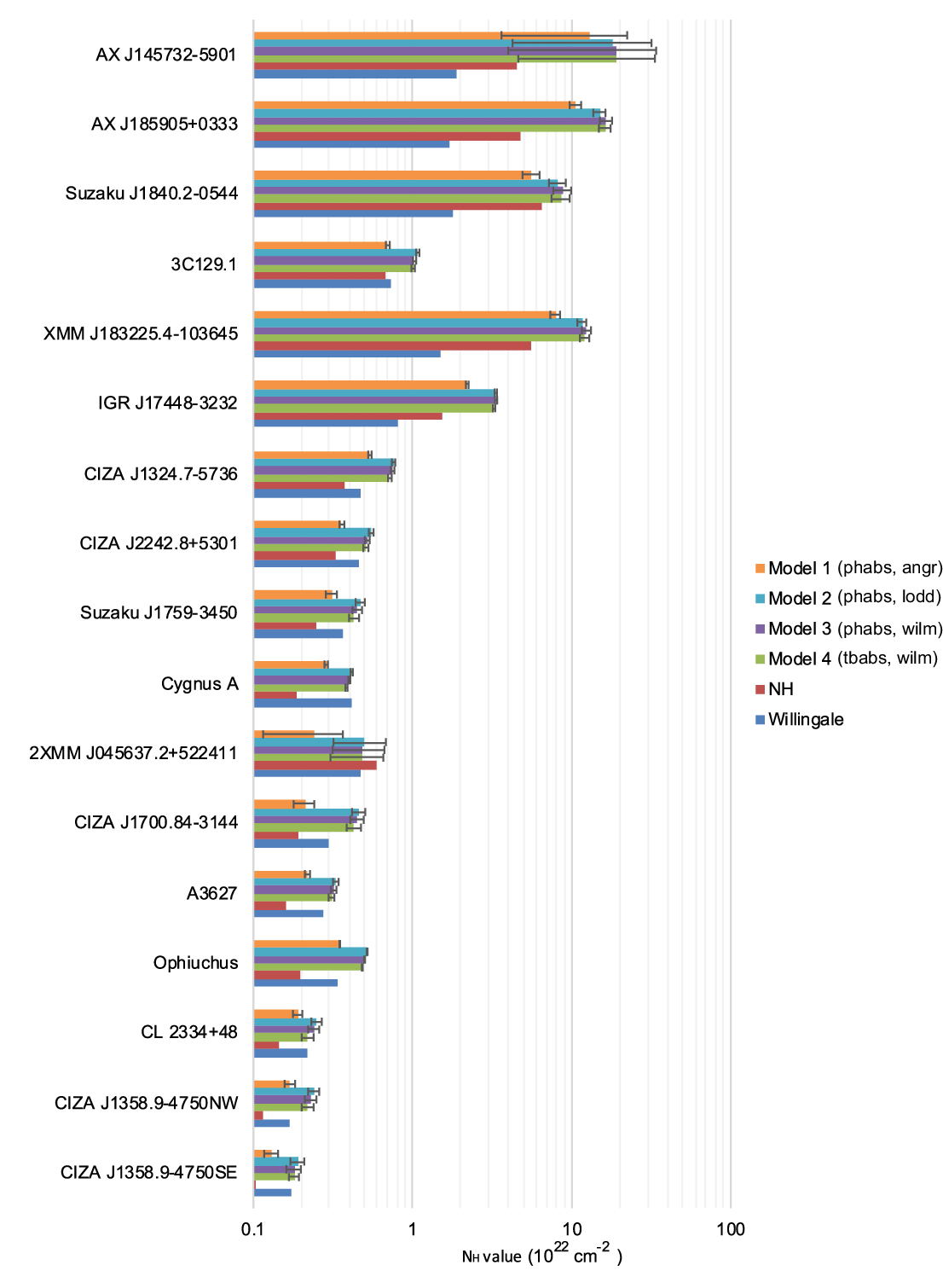}
  \end{center}
  \caption{Comparison of the observed $N_{\rm H}$ value with those estimated from the previous methods. 
  Model 1--4: observed $N_{\rm H}$ values with an error of 68\% confidence level, NH: $N_{\rm H}$ values estimated by equation (3), 
  and Willingale: $N_{\rm H}$ values estimated by the method in \citet{Willingale2013}. 
   {Alt text: A bar chart showing the best-fitting results. }
  }
\label{fig:prev}
\end{figure*}

\subsection{$N_{\rm H}$ values estimated from the previous methods}

Using the results derived from radio band observations, we estimate the Galactic $N_{\rm H_I}$ and $N_{\rm H_2}$ column densities 
of the line of sight toward clusters of galaxies and then evaluate the total $N_{\rm H}$ values by the equation (3). 
The $N_{\rm H_I}$ value was taken from the data in the HI4PI survey \citep{HI4PI2016}. 
The pixel size of the data is \timeform{0.D083}.  
Using web tool\footnote{https://heasarc.gsfc.nasa.gov/cgi-bin/Tools/w3nh/w3nh.pl}, we estimated 
a weighted mean value within a cone radius of \timeform{0.D1}. 
The CO intensity, $W_{\rm CO}$, at the target position was taken from data in \citet{Dame2001}. 
The pixel size is \timeform{0.D125}. We utilized a value in the pixel including the target position. 
To convert $W_{\rm CO}$ to  $N_{\rm H_2}$ value, 
we assumed $X_{\rm CO}$=1.8$\times$10$^{20}$ cm$^{-2}$ K$^{-1}$ km$^{-1}$ s \citep{Dame2001}. 
Then, we obtained the total $N_{\rm H}$ value from the equation (3). 
We also calculated the $N_{\rm H}$ value for the target position 
using the method in \citet{Willingale2013}\footnote{https://www.swift.ac.uk/analysis/nhtot/index.php}.

Figure \ref{fig:prev} displays comparisons of the observed $N_{\rm H}$ value listed in tables \ref{tab:fit1}-\ref{tab:fit3} and the calculated values 
using the previous methods. 
In all cases, the observed $N_{\rm H}$ values are larger than those calculated by equation (3). 
On the other hand, the values by the Willingale method are comparable to those of CGs 
at high Galactic latitude (|$b$| $\gtrsim 5^{\circ}$), 
suggesting that the estimation seems to be 
good, but they are smaller than those of 6 CGs with $N_{\rm H}\gtrsim10^{22}$ cm$^{-2}$ located near to the Galactic plane (|$b$| $< 2^{\circ}$).

\subsection{Correlation between the $N_{\rm H}$ values and $\tau_{353}$}

\begin{figure*}
  \begin{center}
    \FigureFile(8cm,8cm){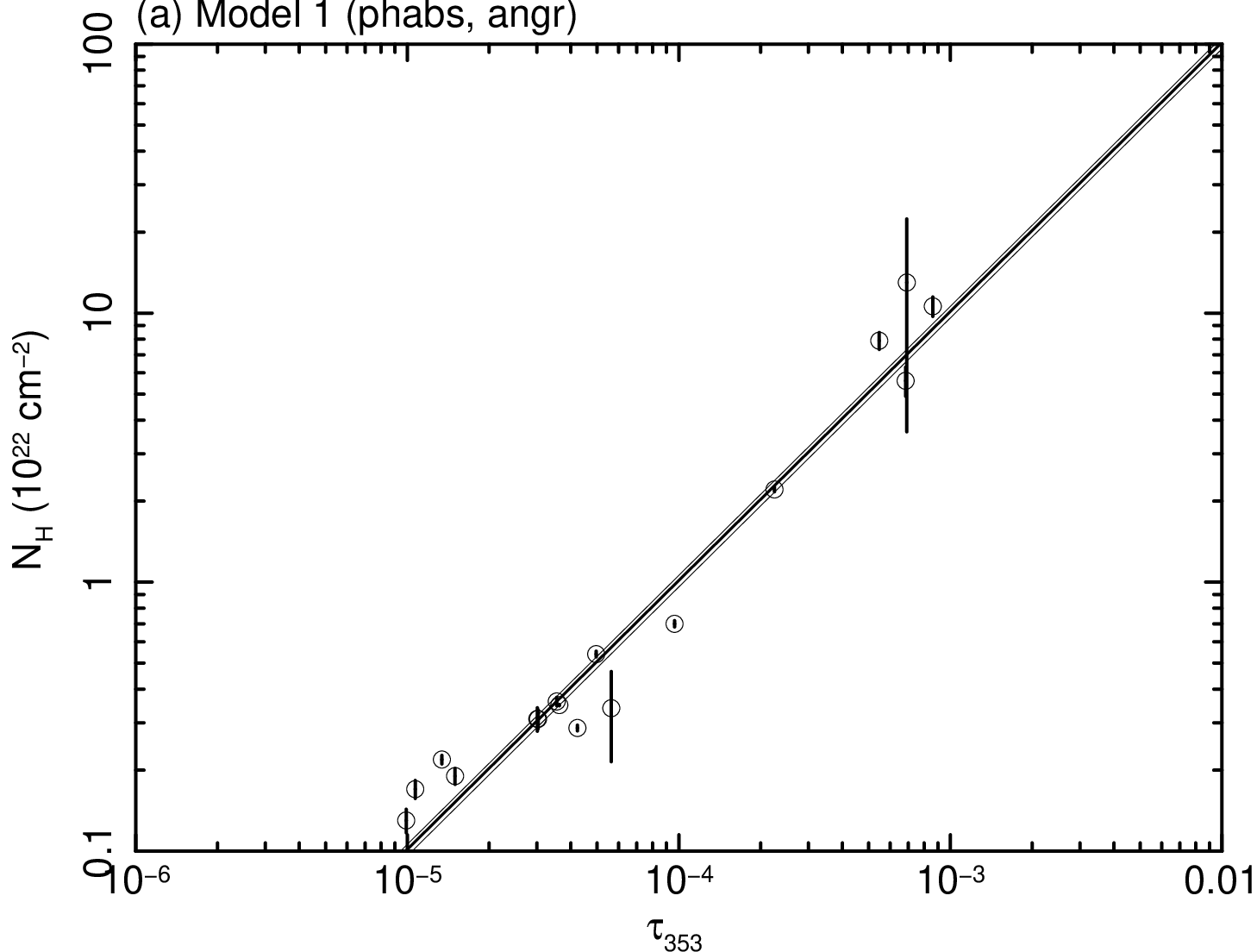}
    \FigureFile(8cm,8cm){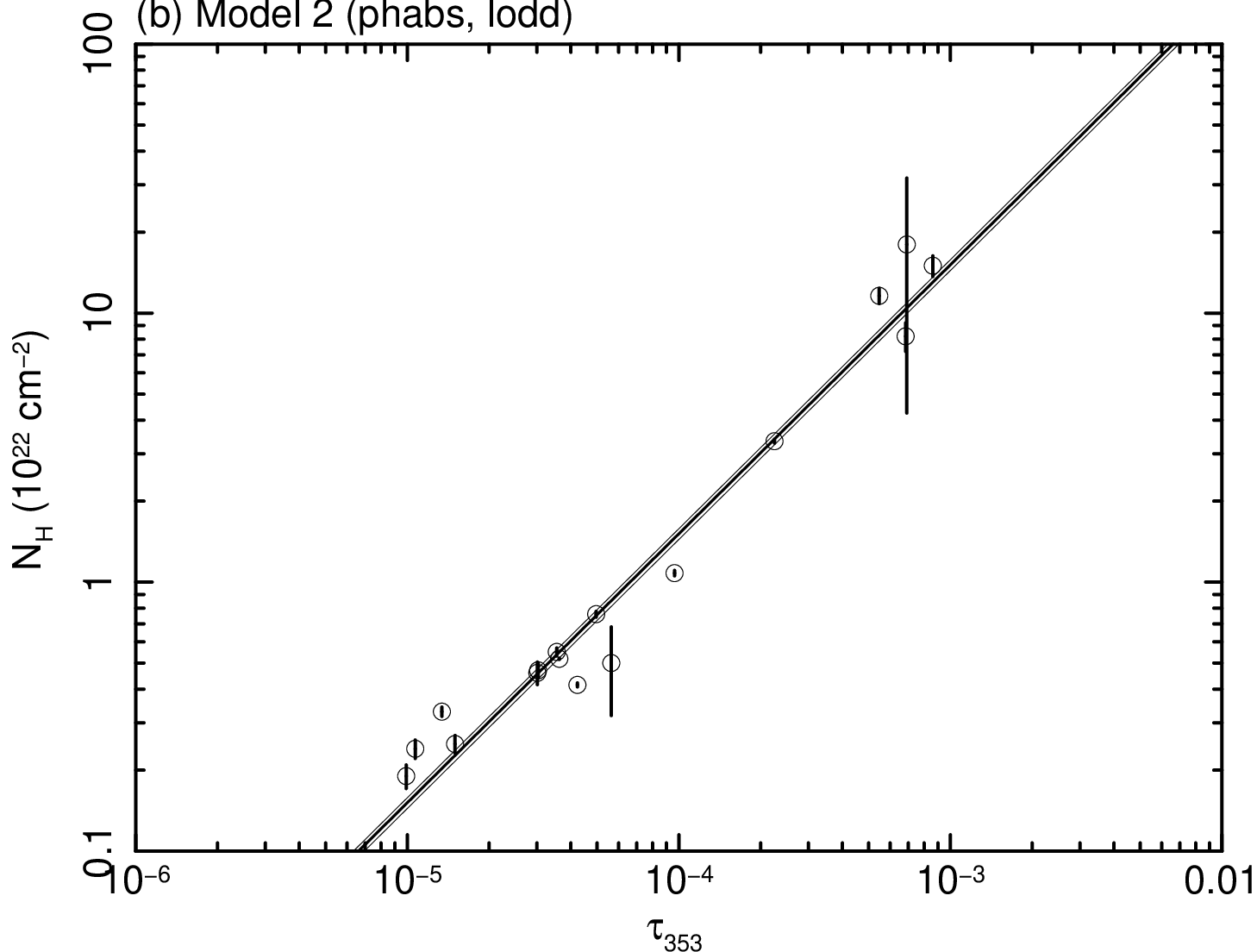}
    \FigureFile(8cm,8cm){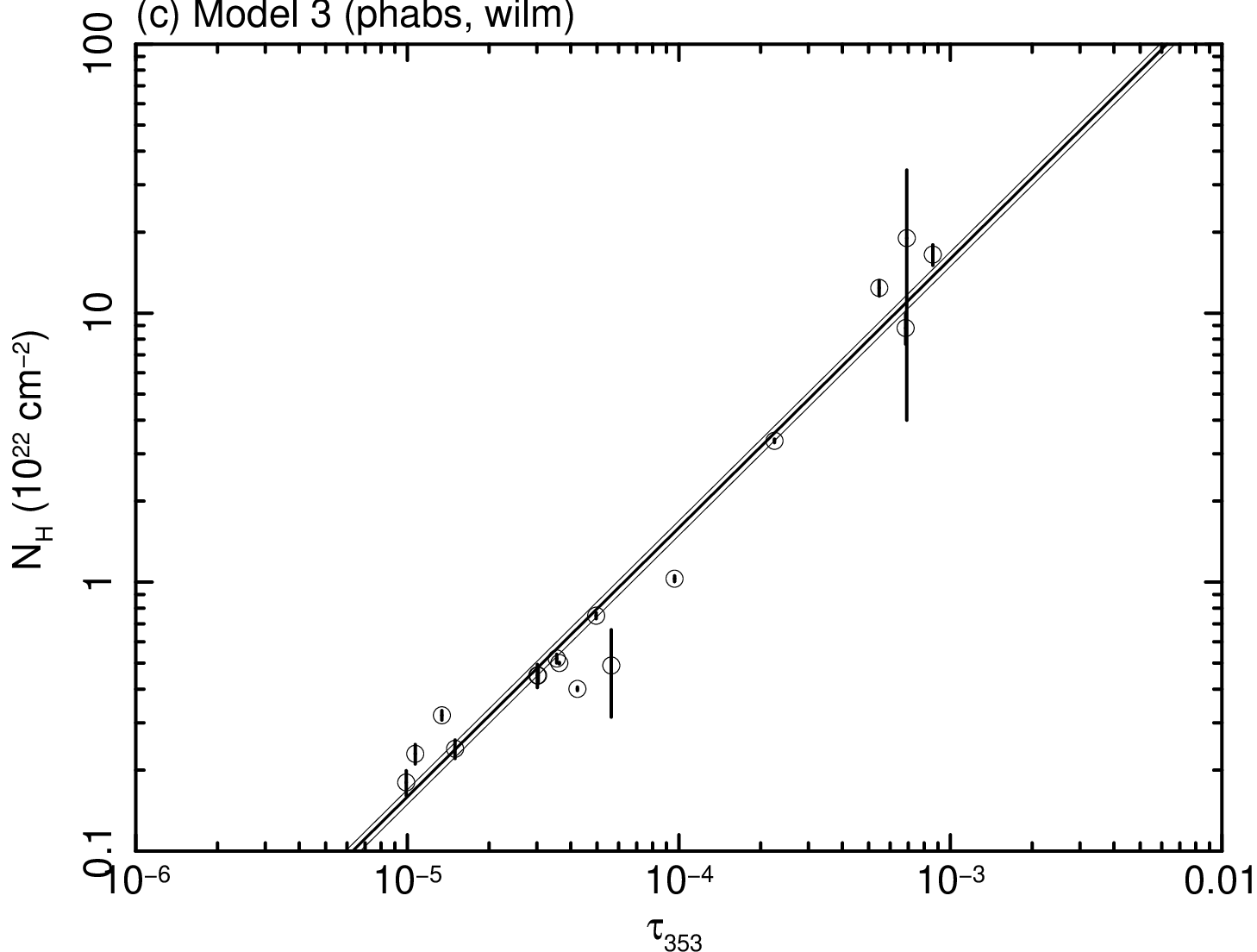}
    \FigureFile(8cm,8cm){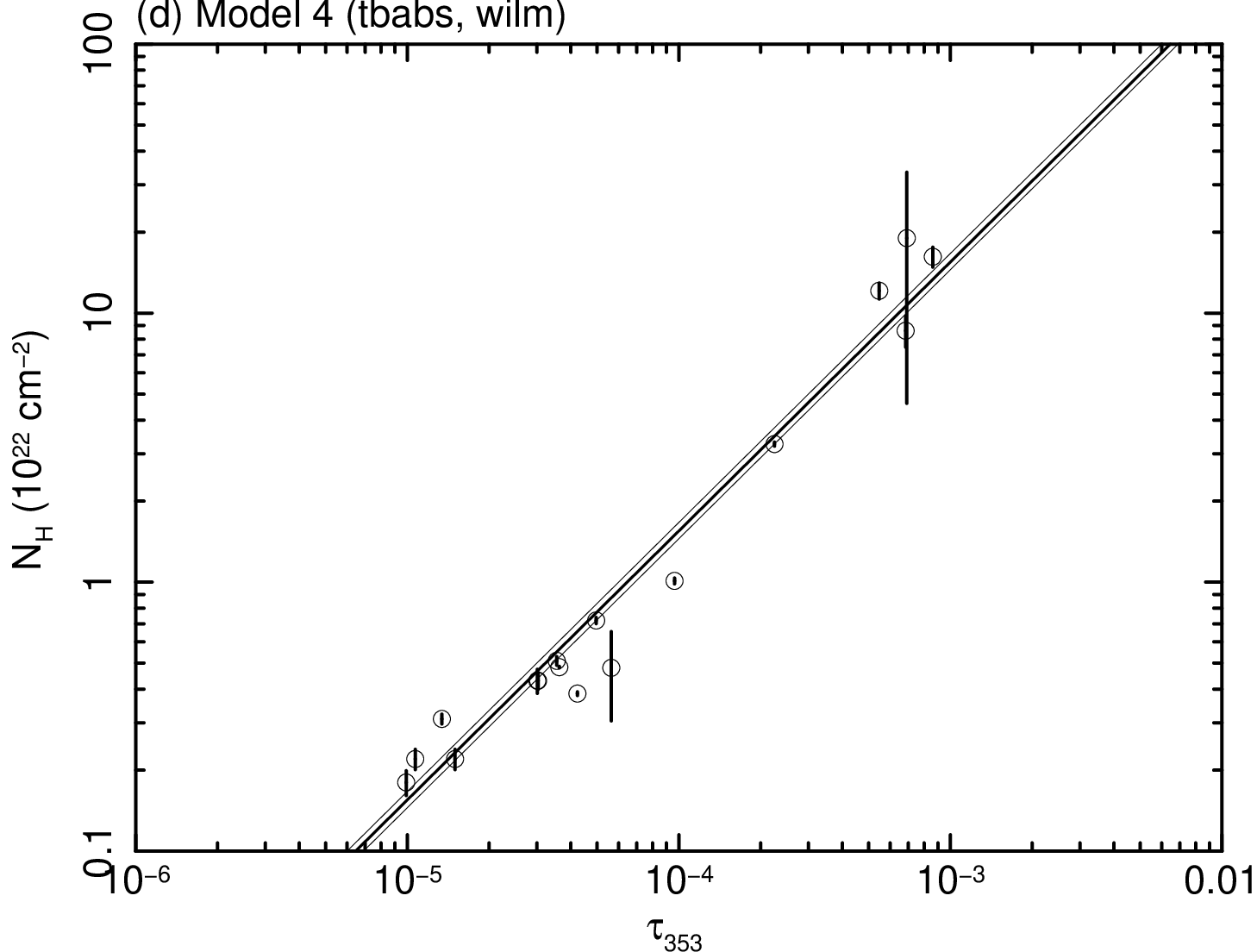}
  \end{center}
  \caption{
  Correlation plot between the optical depth at 353 GHz, $\tau_{353}$, and the observed $N_{\rm H}$ values, 
  (a) Model 1 {\tt (phabs, angr)}, (b) Model 2 {\tt (phabs, lodd)}, (c) Model 3 {\tt (phabs, wilm)}, and (d) Model 4 {\tt (tbabs, wilm)}.
  Vertical errors show a 68\% confidence level. 
 The thick line shows the best-fitting model (see equation 5--8), whereas the thin lines show the model curve 
 obtained by adding $\pm 1 \sigma$ errors of the coefficients of the linear function. 
 {Alt text: Line graphs showing the best-fitting results. }
  }
\label{fig:plot}
\end{figure*}

The optical depth at 353 GHz, $\tau_{353}$, is much less than 1, 
and hence the values of $\tau_{353}$ 
can trace
the amount of matter along the line of sight. 
Using $\tau_{353}$ from the Planck all-sky data\footnote{http://irsa.ipac.caltech.edu/data/Planck/release\_1/all-sky-maps/}, 
we examined a correlation between the $\tau_{353}$ and the $N_{\rm H}$ values. 
The pixel size of the $\tau_{353}$ data is \timeform{1'.7}. 
We used a mean $\tau_{353}$ value of $3\times3$ pixels centered on the pixel including the target position. 
Figure \ref{fig:plot} shows a correlation plot between the $\tau_{353}$ and the $N_{\rm H}$ values derived from Model 1--4. 
All the models show that the observed $N_{\rm H}$ values well correlates with the $\tau_{353}$ values.  
The correlation coefficients are calculated to be 0.94--0.95. 

We fitted the data with a linear relation using the Markov chain Monte Carlo (MCMC) method implemented in the Python package {\tt emcee} 
\citep{Foreman-mackey2013}. 
The measurement uncertainties were taken into account, and the intrinsic scatter was treated as a free parameter. 
The slope was derived from the posterior distribution as follows: 
\begin{equation}
{\rm Model\ 1}\ {\tt (phabs, angr)}: N_{\rm H}=(1.01\pm0.05)\times10^{26}\ \tau_{353}\ {\rm cm}^{-2}\\
\end{equation}
\begin{equation}
{\rm Model\ 2}\ {\tt (phabs, lodd)}: N_{\rm H}=(1.51\pm0.07)\times10^{26}\ \tau_{353}\ {\rm cm}^{-2}\\ 
\end{equation}
\begin{equation}
{\rm Model\ 3}\ {\tt (phabs, wilm)}: N_{\rm H}=(1.59\pm0.10)\times10^{26}\ \tau_{353}\ {\rm cm}^{-2}\\
\end{equation}
\begin{equation}
{\rm Model\ 4}\ {\tt (tbabs, wilm)}: N_{\rm H}=(1.55\pm0.10)\times10^{26}\ \tau_{353}\ {\rm cm}^{-2}\\ 
\end{equation}
The errors are statistical errors with a 68\% confidence level. 
The best-fitting model is displayed in each panel.

\subsection{Correlation between the $N_{\rm H}$ values and $E(B-V)$}

\begin{figure*}
  \begin{center}
    \FigureFile(8cm,8cm){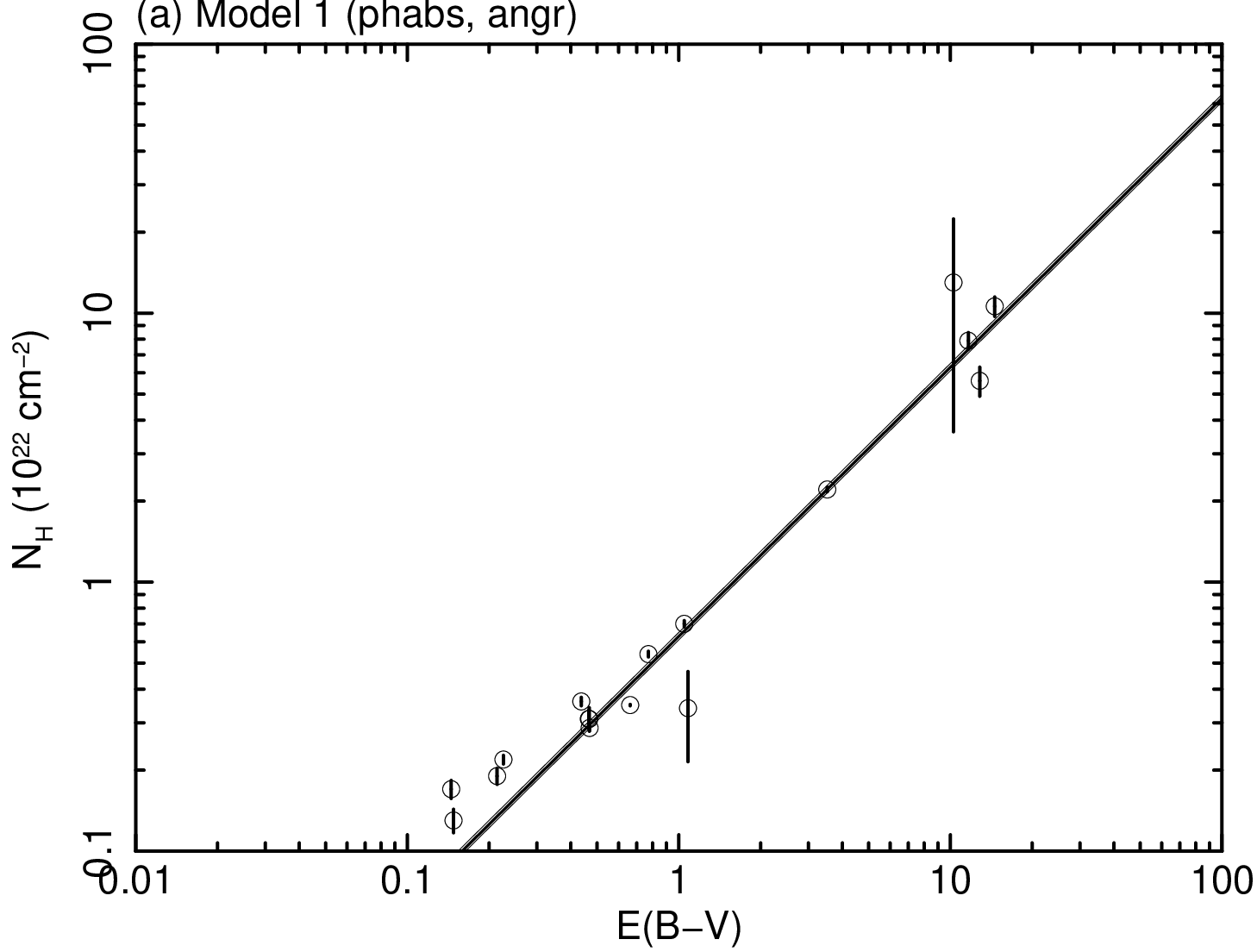}
    \FigureFile(8cm,8cm){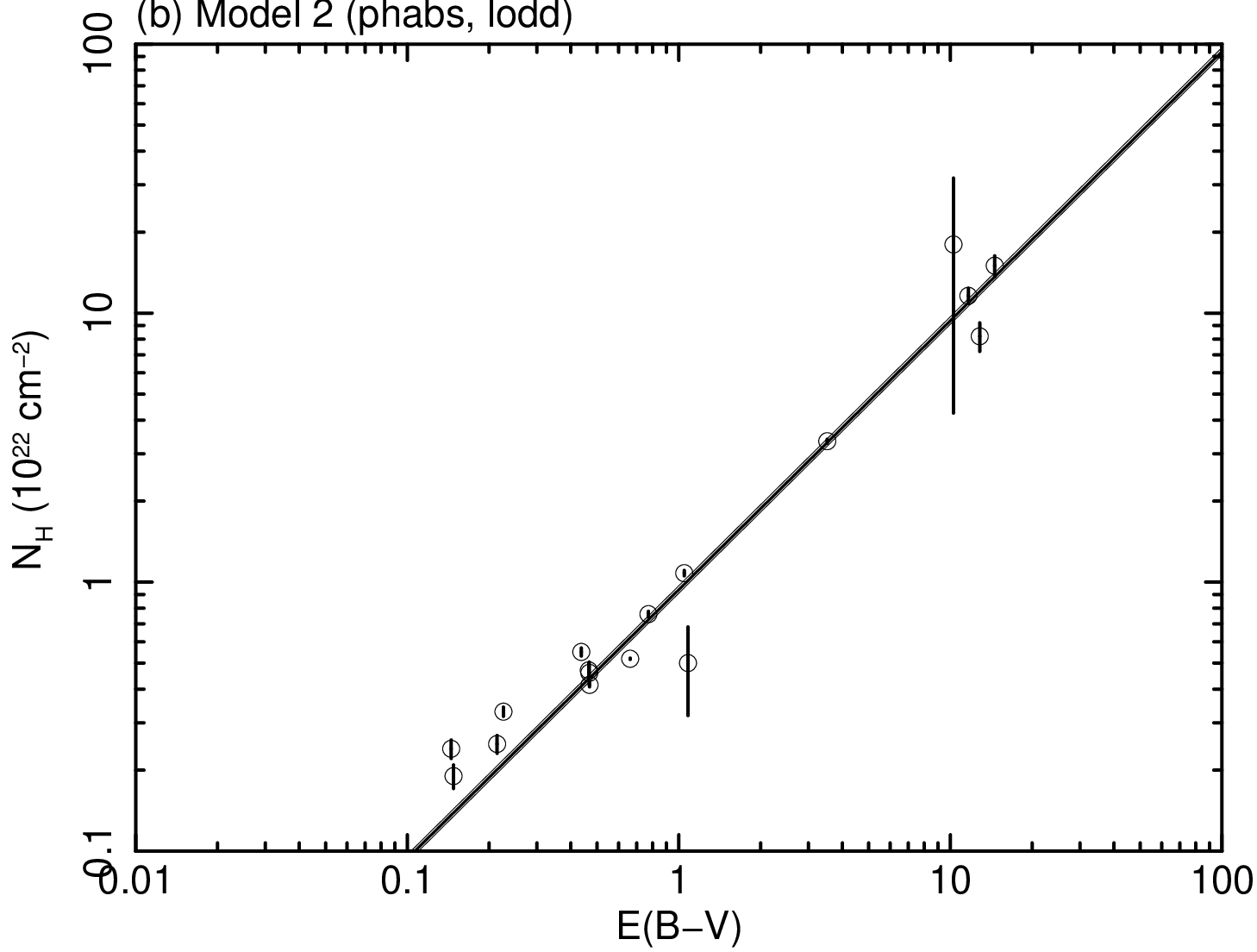}
    \FigureFile(8cm,8cm){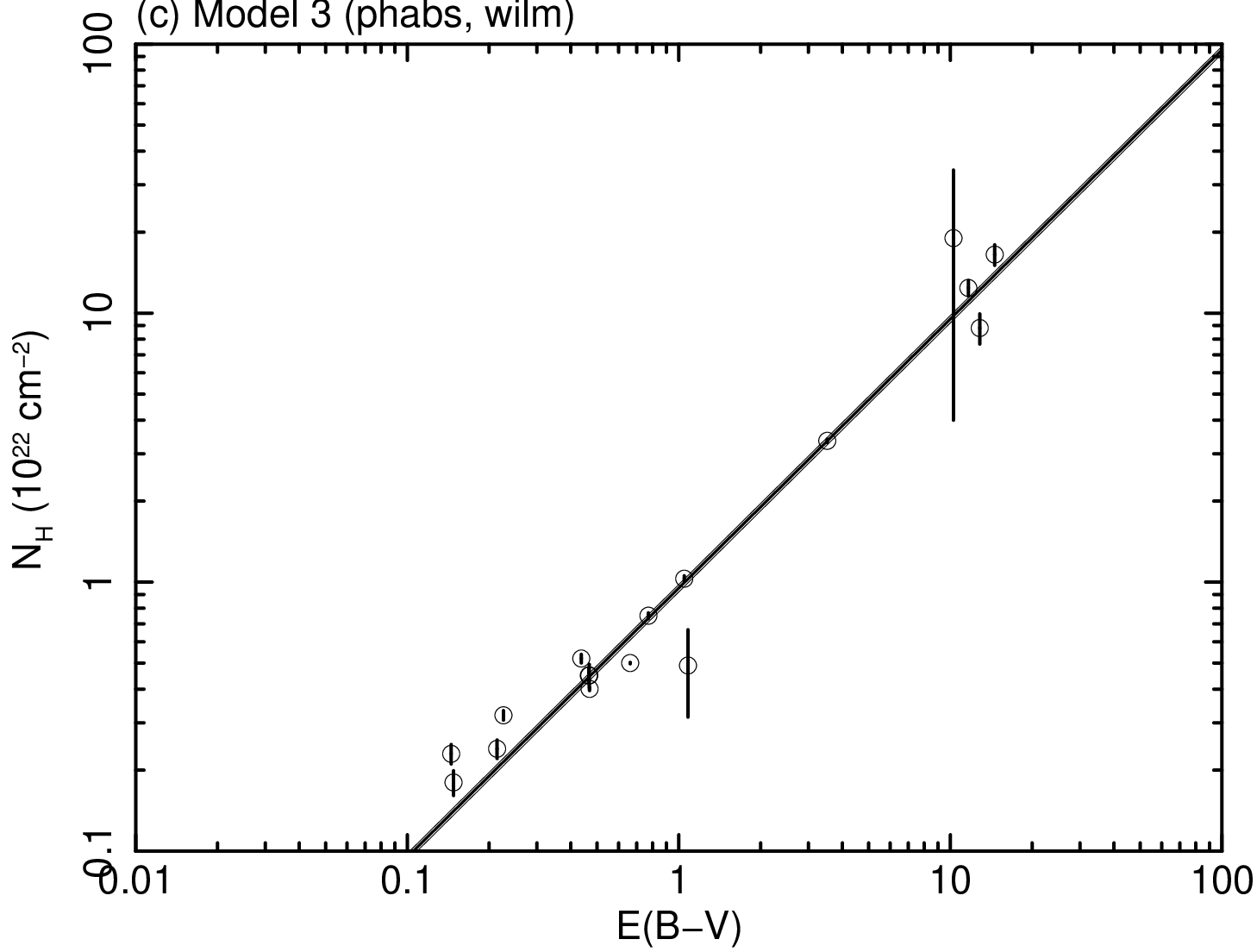}
    \FigureFile(8cm,8cm){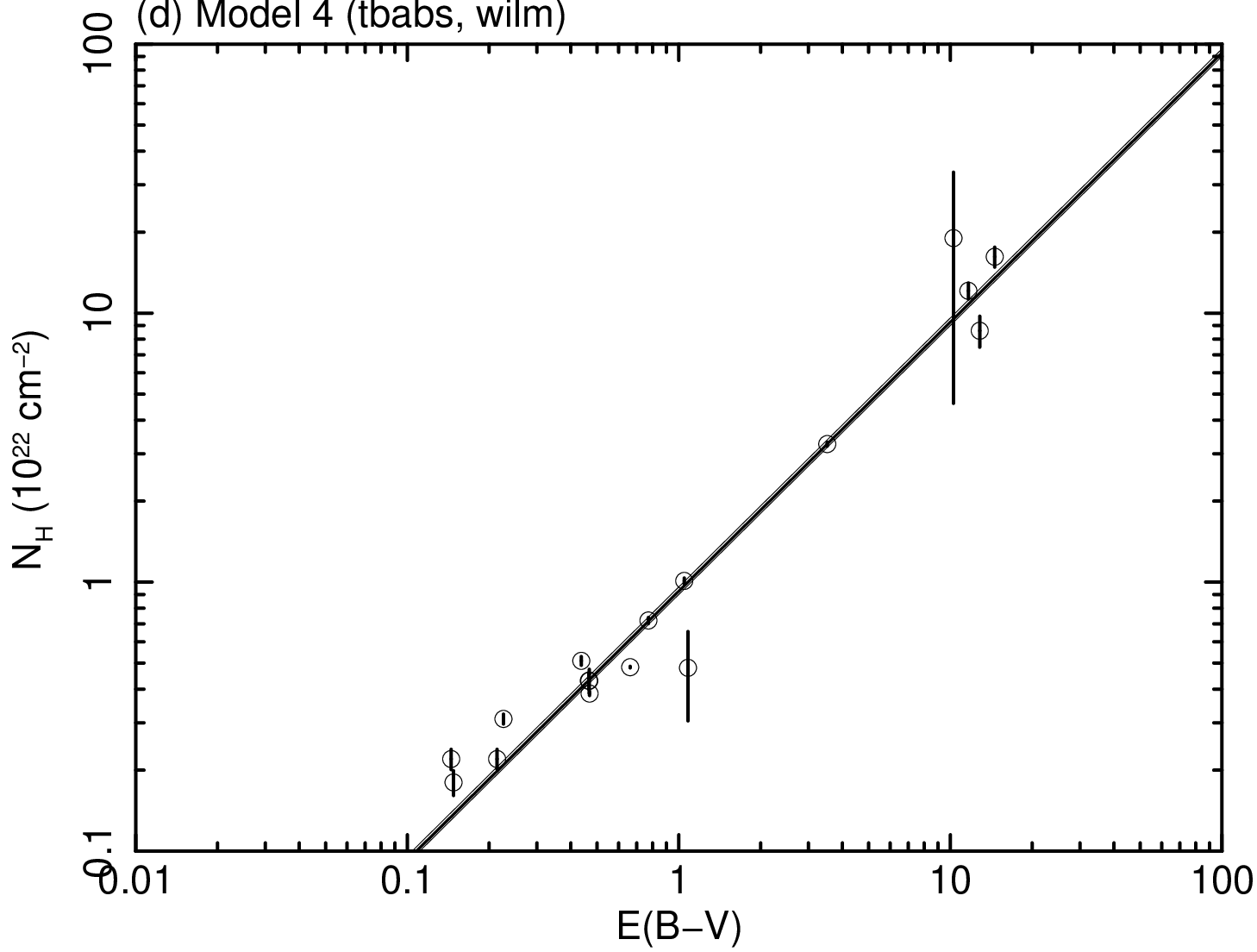}
  \end{center}
  \caption{
  Correlation plot between the optical reddening, $E(B-V)$, and the observed $N_{\rm H}$ values, 
    (a) Model 1 {\tt (phabs, angr)}, (b) Model 2 {\tt (phabs, lodd)}, (c) Model 3 {\tt (phabs, wilm)}, and (d) Model 4 {\tt (tbabs, wilm)}.
  Vertical errors show a 68\% confidence level. 
 The thick line shows the best-fitting model (see equation 9--12), whereas the thin lines show the model curve 
 obtained by adding $\pm 1 \sigma$ errors of the coefficients of the linear function. 
 {Alt text: Line graphs showing the best-fitting results. }
  }
\label{fig:plot2}
\end{figure*}

The correlation with between $E(B-V)$ and $N_{\rm H}$ values is well known facts as described in section 1. 
Thus, we also examined the relationship between $E(B-V)$ and $N_{\rm H}$ values obtained in the present work. 
The $E(B-V)$ values were referred to the data by \citet{Schlegel1998}\footnote{Taken from https://www.swift.ac.uk/analysis/nhtot/index.php.}. 
Figure \ref{fig:plot2} shows a correlation plot between the $E(B-V)$ and the $N_{\rm H}$ values derived from Model 1--4. 
As same as the $\tau_{353}-N_{\rm H}$ relation, we fitted the data with a linear function and obtained the following results. 
\begin{equation}
{\rm Model\ 1}\ {\tt (phabs, angr)}: N_{\rm H}=(6.3\pm0.2)\times10^{21}\ E(B-V)\ {\rm cm}^{-2}\\
\end{equation}
\begin{equation}
{\rm Model\ 2}\ {\tt (phabs, lodd)}: N_{\rm H}=(9.4\pm0.3)\times10^{21}\ E(B-V)\ {\rm cm}^{-2}\\ 
\end{equation}
\begin{equation}
{\rm Model\ 3}\ {\tt (phabs, wilm)}: N_{\rm H}=(9.5\pm0.3)\times10^{21}\ E(B-V)\ {\rm cm}^{-2}\\
\end{equation}
\begin{equation}
{\rm Model\ 4}\ {\tt (tbabs,  wilm)}: N_{\rm H}=(9.3\pm0.3)\times10^{21}\ E(B-V)\ {\rm cm}^{-2} \\ 
\end{equation}
The errors are statistical errors with a 68\% confidence level. 
The best-fitting model is displayed in each panel.

\section{Discussion and conclusion}

We analyzed 17 spectra of 16 clusters of galaxies and the candidates located at the low Galactic latitude 
using three abundance tables and two photoelectric absorption models,  
and obtained $N_{\rm H}$ values along the line of sight. 
We found that the $N_{\rm H}$ values using abundance tables of {\tt lodd} and {\tt wilm} are consistent with each other, 
but the values of  {\tt angr} are about 1.6 times smaller than those of {\tt lodd} and {\tt wilm}. 
This is because that absorption by O is large in the low energy X-rays below 1 keV and 
the abundance of O relative to H in {\tt angr} is 1.7 times larger than those in {\tt lodd} and {\tt wilm}. 
On the other hand, 
differences in the cross sections of photoelectric absorption models of {\tt phabs} and {\tt tbabs} are negligible. 

We also note that the metal abundances of the X-ray emitting plasma in clusters of galaxies derived using {\tt lodd} and {\tt wilm} tables are 1.6--1.7 times larger than that from {\tt angr}. 
The metal abundances are estimated from intensities of emission lines in the spectrum. 
The intensity of the Fe-K line is a good tracer in the plasma with a temperature of $\sim$3--10 keV, thus 
the difference in the metal abundance in fitting with three abundance tables is because that the abundance of Fe relative to H in {\tt angr} is larger than those in {\tt lodd} and {\tt wilm}. 

We found that in all the cases of Model 1--4, the observed $N_{\rm H}$ values are systematically larger than 
those calculated from equation (3), 
in agreement with the fact that a considerable amount of gas is not traced by standard H\emissiontype{I} or 
CO line surveys \citep{Grenier2005}.
On the other hand, the observed values at high Galactic latitude 
($N_{\rm H}$ $\lesssim 10^{22}$ cm$^{-2}$) are comparable to values estimated by equation (4), 
but the values near to the Galactic plane ($N_{\rm H} >10^{22}$ cm$^{-2}$) are larger than the estimated values.

Fukui et al. (2014, 2015) pointed out that the relation of 
the H\emissiontype{I} integrated intensity and the optical depth at 353 GHz depends on the dust temperature and  
saturation of the H\emissiontype{I} emission line at the 21 cm wavelength occurred in the case of low dust temperature.  
They proposed that optically thick H\emissiontype{I} would be a likely scenario. 
Since the values of optical depth of dust at 353 GHz are in the range of $10^{-5}$--$10^{-3}$, 
dust radiation is completely transparent through the Galaxy. 
\citet{Fukui2014} showed a relationship between $N_{\rm H\emissiontype{I}}$ and $\tau_{353}$ of $N_{\rm H\emissiontype{I}}=1.5\times10^{26}$\ $\tau_{353}$ cm$^{-2}$ 
in the optically thin assumption with a high dust temperature. 

The X-ray absorption process is the photoelectric absorption, 
which depends on the number of atoms to encounter X-ray photons, 
and hence  
X-ray observations would be able to derive the $N_{\rm H}$ values independently of the condition of the ISM. 
Our results show that the obtained $N_{\rm H}$ values in the range of $10^{21}$--$10^{23}$ cm$^{-2}$ 
are well correlated with the optical depth at 353 GHz, $\tau_{353}$ (see figure \ref{fig:plot}). 
The dust temperatures of our samples are in the range of 16.8--22.1 K, but the correlation shows no dependence on the dust temperature.
The coefficients depend on the abundance table: the value from {\tt angr} is smaller than those from  {\tt wilm} and {\tt lodd}. 
The relations in the cases of {\tt wilm} and {\tt lodd}, relatively newer abundance tables, 
are $N_{\rm H} \sim 1.5\times10^{26}\ \tau_{353}$ cm$^{-2}$, 
which is in agreement with the $N_{\rm H\emissiontype{I}}$--$\tau_{353}$ relation in \citet{Fukui2014}. 

We also
confirmed that  $E(B-V)$ and the observed $N_{\rm H}$ values are linearly correlated even at $E(B-V)>10$ mag and $N_{\rm H}>10^{23}$ cm$^{-2}$, 
expressed by $N_{\rm H}=(6.3-9.5)\times10^{21}\ E(B-V)\ {\rm cm}^{-2}$. 
These are consistent with those obtained in the previous works \citep{Bohlin1978,Hensley2017, Lenz2017,Li2018}. 

In this work, we assumed a single $N_{\rm H}$ value in the spectral analysis for diffuse X-ray emission of clusters of galaxies. 
\citet{Locatelli2022} states that if the spatial distribution of $N_{\rm H}$ is not considered, X-ray spectral analysis can yield values smaller than 
the actual $N_{\rm H}$ value and 
the impact on the value is several times greater, especially when $N_{\rm H}>10^{22}$ cm$^{-2}$. 
This would cause deviations from the linear relationships of $\tau_{353}-N_{\rm H}$ and  $E(B-V)-N_{\rm H}$: 
the data with the large $N_{\rm H}$ would take values smaller than the linear relationship. 
The smaller variance of $N_{\rm H}$ value within the source region may give accurate $N_{\rm H}$ estimation. 
Detailed analysis taking account of the spatial distribution of the ISM column density within the source region would provide more accurate estimation. 

Assuming that the plasma of clusters of galaxies is isothermal, we fitted the spectra of clusters of galaxies with a single temperature component model. 
Since some clusters of galaxies exhibit temperature distribution, such as cool core and low temperature outskirt, the obtained $N_{\rm H}$ values may have some difference. 
These may cause the variance around the fitted line. 
This could be verified by examining the spatial distribution of the plasma temperature within the source region. 

\section*{Funding}
This work was supported by the Japan Society for the Promotion of Science (JSPS) KAKENHI Grant Numbers 21K03615 and 
24K00677 (SY, KKN, MN, and HU).

\ack
This research
made use of the NASA/IPAC Extragalactic Database (NED) operated by Jet Propulsion Laboratory, California Institute of Technology, 
under contract with NASA.


\end{document}